\documentclass[aps,twocolumn,floatfix,showpacs]{revtex4}
\usepackage{graphicx}
\usepackage{epsfig}

\def \tcomp {T_{\rm comp}}
\def \ncomp {n_{\rm comp}}
\def \TC {T_{\rm c}}
\def \ld {{L^{{\rm cf}}}}
\def \lo {\lambda }
\def \vS {{\bf S}}
\def \vL {{\bf L}}
\def \vJ {{\bf J}}

\begin{document}

\title{Hybrid Quantum-Classical Monte-Carlo Study of a Molecule-Based Magnet}

\author{P. Henelius}
\affiliation{Theoretical Physics, Royal Institute of Technology, SE-106 91 Stockholm, Sweden}

\author{R.S. Fishman}
\affiliation{Materials Science and Technology Division, Oak Ridge National Laboratory, Oak Ridge, Tennessee 37831-6065, USA}

\date{\today}

\begin{abstract}

Using a Monte Carlo (MC) method, we study an effective model for the Fe(II)Fe(III) bimetallic oxalates. 
Within a hybrid quantum-classical MC algorithm, the Heisenberg $S=2$ and $S'=5/2$ spins on the 
Fe(II) and Fe(III) sites are updated using a quantum MC loop while the Ising-like orbital angular momenta on 
the Fe(II) sites are updated using a single-spin classical MC flip. The effective field acting on the orbital 
angular momenta depends on the quantum state of the system.  We find that the mean-field phase diagram for the model  
is surprisingly robust with respect to fluctuations.  In particular, the region displaying two compensation points shifts and 
shrinks but remains finite.

\end{abstract}

\maketitle

\section{Introduction}

\begin{figure}
\resizebox{\hsize}{!}{\includegraphics{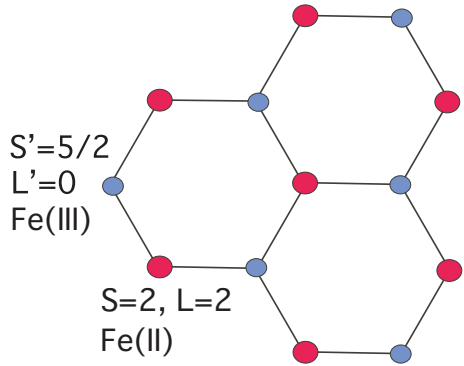}}
\caption{The honeycomb lattice showing the alternating Fe(II) and Fe(III) sites.}
\label{honey} 
\end{figure}

Bimetallic oxalates are layered, molecule-based magnets with the chemical formula 
A[M(II)M'(III)(ox)$_3$] \cite{Tamaki92}.  Every layer contains two different 
transition metal atoms, M(II) and M'(III), in the alternating honeycomb structure depicted in Fig.\ref{honey}.
Each bond represents an oxalate molecule ox = C$_2$O$_4$, which generates 
a crystal-field potential at both ionic sites.  For different transition metals, 
bimetallic oxalates can be ferromagnetic, antiferromagnetic, or ferrimagnetic with moments always
pointing out of the plane \cite{Clem03}.  Since the type of magnetic order does not depend 
on the cation A that couples the magnetic layers, the magnetic properties of the bimetallic oxalates 
are primarily controlled by a single bimetallic layer.  

For the Fe(II)Fe(III) bimetallic oxalates, however, the presence of magnetic compensation 
below the ferrimagnetic transition temperature $\TC $ does depend 
on the choice of  A \cite{comp}.  For several cations, the magnetization in a small field
is positive just below $\TC \approx 45$ K but then become negative below $\tcomp \approx 32$ K.
This effect was explained by Fishman and Reboredo \cite{Fish07}, who used mean-field (MF) theory to
solve an effective Hamiltonian that includes spin-orbit coupling on the Fe(II) sites.  Magnetic compensation is
produced when the orbital angular momentum $\ld $ of the low-lying crystal-field doublet on the Fe(II) sites exceeds 
a threshold value.  By altering the crystal-field potential, the cation A can shift $\ld $ above or below this threshold value.  
In this paper, we use a Monte-Carlo (MC) technique to study the same effective Hamiltonian and demonstrate that the 
MF results are surprisingly immune to the effect of fluctuations.

By Hund's first rule, the spin on the Fe(II) (3d$^6$)and Fe(III) (3d$^5$) sites are $S=2$ and $S'=5/2$, respectively.
Since the Fe(III) multiplet is half-full, its orbital angular momentum $L'$ vanishes according to Hund's second rule.  
The $L=2$ orbital angular momentum on the Fe(II) sites is split by the $C_3$-symmetric crystal-field potential 
produced by the 6 oxygen atoms surrounding each ion.  This splitting creates two doublets and one singlet \cite{Fish07}.  
The orbital angular momentum is unquenched when one of the doublets lies lowest in energy.  
In that case, the out-of-plane or $z$ component of the 
orbital angular momentum on the Fe(II) sites takes values $\pm \ld $, where $\ld $ ranges from 
0 to 2 and depends on the crystal-field potential.  The spin-orbit coupling on the Fe(II) sites is given by 
$\lo \vL_i\cdot \vS_i$, where $\vL_i=\pm \ld {\bf z}$ and 
$\lo \approx -147$ K \cite{Bleaney53} is the spin-orbit coupling constant (negative because the 3d$^6$ shell
is more than half-filled).  The total angular momentum $\vJ_i =\vS_i +\vL_i $ on the Fe(II) sites
is not a good quantum number and Hund's third rule is not obeyed because the crystal-field 
potential is large compared with the spin-orbit coupling.

Hence, the effective Hamiltonian of the Fe(II)Fe(III) bimetallic oxalates can be written
\begin{equation} 
\label{ham}
H= J\sum_{\langle ij\rangle } \vS_i\cdot \vS'_j +\lambda \sum_i  S^z_i L^z_i, \label{ls}
\end{equation}
where the $\langle ij \rangle $ summation in the exchange term 
is performed over all nearest neighbors on the honeycomb lattice and the 
$i$ summation in the spin-orbit term is performed over the Fe(II) sites only. 
The orbital angular momentum $L^z_i=\pm \ld $ can be treated as a classical variable so long as the 
relevant energy scales are smaller than the splitting $\Delta $ between the lowest-energy doublet and the nearest
excited states of the crystal-field potential.   Since $\TC $ is less than 45 K while $\Delta $ is 
larger than room temperature, this should be a good assumption for the bimetallic oxalates.  For 
the special case where the singlet lies lowest in energy, we would take $\ld =0$.

Several other models can also explain the existence of magnetic compensation in the Fe(II)Fe(III) bimetallic oxalates.
Nakamura \cite{Nak00} described the Fe(II) and Fe(III) spins by $S=2$ and $S'=5/2$ Ising variables with 
single-ion anisotropy $-D(S_i^z)^2$ on the $S=2$ sites.  Although magnetic compensation is found 
in a MF or effective-field theory treatment of this model, it is absent when the model is solved using the MC method.  
With fluctuations correctly included, magnetic compensation is recovered only after interlayer or longer-ranged 
interactions are considered.  A similar conclusion was reached by Carling and Day \cite{Carling01}, 
who used MC simulations to show that next-neighbor interactions between Ising spins on the same 
sublattice are required to obtain magnetic compensation within a single bimetallic layer.  
Li {\em et al.} \cite{Li04} treated the spins as Heisenberg operators 
with interlayer interactions and single-ion anisotropy on both sublattices.  Using a 
Green's function decoupling scheme, they found that magnetic compensation occurs even in the absence of 
interlayer interactions when anisotropy is experienced by a single magnetic sublattice.   
However, it is uncertain whether magnetic compensation would survive in the presence of correlated thermal fluctuations.

Those other treatments fail to explain several important features of the Fe(II)Fe(III) bimetallic oxalates that can be
explained by the model of Eq.(\ref{ham}).  First, Eq.(\ref{ham}) provides a natural explanation for the origin
of the magnetic anisotropy on the Fe(II) sites due to the splitting of the $L=2$ multiplet by the crystal-field potential.
Second, unlike the model studied by Nakamura \cite{Nak00}, the model studied in this paper does not rely on 
interlayer interactions to produce magnetic compensation.  So it can explain the appearance
of magnetic compensation in the A = N(n-C$_n$H$_{2n+1}$)$_4$ family:  as $n$ increases from 3 to 5, 
the interlayer separation grows from 8.2 \AA $\,$ to 10.2 \AA $\,$ \cite{comp} but magnetic compensation 
appears only for $n=4$ and 5, probably because defects in the bimetallic layer are created by the shortest
cation.  Third, Eq.(\ref{ham}) can explain the persistence
of negative magnetization below $\tcomp $ in small fields:  flipping the orbital angular momentum $L_i^z$
requires an energy of $2\vert \lo \vert \ld S \approx 175$ K $\gg \tcomp $.  By contrast, the single-ion anisotropy 
$D \sim \lo^2 /\Delta $ experienced by the Fe(II) moments in most materials is less than 10 K \cite{Rud03}.
So the energy barrier $4D$ for flipping the Fe(II) spin is smaller than about 40 K, which is comparable to 
$\tcomp \approx 32$ K.  Finally, Eq.(\ref{ham}) explains the recently-observed jump \cite{Tang07} in the 
magnetization between $\tcomp $ and $\TC $, which is believed to arise from an inverse Jahn-Teller transition \cite{Fish08}.
Indeed, it is difficult to explain this jump based on the other models introduced above.

Magnetic compensation occurs in a MF treatment \cite{Fish07} of Eq.(\ref{ham}) for two reasons:  
the anisotropy at the Fe(II) sites produced by spin-orbit coupling and the Fe(II)
orbital contribution to the total magnetic moment.  The latter is 
absent in an Ising or Heisenberg ferromagnet with single-ion anisotropy.  While MF theory was 
used to approximate the exchange coupling $J \vS_i \cdot \vS'_j $ between neighboring 
Fe(II) and Fe(III) moments, the spin-orbit coupling $\lo \vL_i \cdot \vS_i $ at the Fe(II) sites was 
treated exactly within the crystal-field doublet.  As demonstrated by Nakamura \cite{Nak00}, however,
models that exhibit magnetic compensation within MF or effective-field theories may no longer do so 
once fluctuations are correctly included.  So a MC study including both quantum and thermal fluctuations 
is needed to confirm that Eq.(\ref{ham}) supports magnetic compensation.  

This paper is divided into four sections.  In the next section, we describe the hybrid MC method used to study
Eq.(\ref{ham}).  The results of our MC study and a comparison with earlier MF results
are presented in Section III.  A brief conclusion is provided in Section IV.

\section{Hybrid MC Method}

We apply the stochastic series expansion (SSE) quantum MC method \cite{Sand99} to the model Hamiltonian $H$. 
The SSE method employs a Taylor expansion of the partition function $Z$:
\begin{equation}
Z=\sum_{\alpha}\sum_{n=0}^{\infty} \frac{(-\beta)^n}{n!}\langle\alpha
| H^n |\alpha\rangle,
\end{equation}
where $|\alpha\rangle$ are the basis states used to evaluate the matrix elements of $H$
and $\beta =1/T$ is the inverse temperature.
Quantum MC methods have been previously applied to high-spin models and here we follow the 
method described in Ref.\cite{Hen02}.  As explained below, that method has been modified to treat 
the Ising-like orbital angular momentum.  For a more detailed account of the general SSE method, 
we refer to Ref.\cite{Sylj02}. 

To formulate the updating procedure, we write the Hamiltonian as a sum
over all nearest neighbors in the system
\begin{equation}
H=-J\sum_{\langle ij \rangle} H_{ij}.
\end{equation}
The operator $H_{ij}$ can be decomposed into its diagonal and
off-diagonal parts:
\begin{equation}
H_{ij}=H_{D,ij}-H_{O,ij},
\end{equation}
where the subscript $D$ denotes a diagonal operator and $O$ an
off-diagonal operator.  For the present model, these two
operators take the form
\begin{equation}
H_{D,ij}=C-S_i^zS'^z_j + \frac{\lo }{3J}S_i^zL^z_i 
\end{equation}
and
\begin{equation}
H_{O,ij}=\frac{1}{2}\left(S_i^+S'^-_j+S_i^-S'^+_j\right),
\end{equation}
The constant $C$ is included in order to ensure a positive weight in the expansion. 

Introducing a cutoff $K$ in the Taylor expansion (which,
when done properly, does not cause any systematic
errors \cite{Sylj02}) and including additional unit operators $I$, the
expansion can be rewritten as
\begin{equation}
Z=\sum_{\alpha}\sum_{S_K} \frac{\beta^n(K-n)!}{K!}\langle\alpha
| S_K |\alpha\rangle,
\label{taylor}
\end{equation}
where $S_K$ is the operator string
\begin{equation}
S_K=\prod_{p=1}^K H_p,
\end{equation}
with $H_p\in \{H_{D,ij}, H_{O,ij}, I\}$.  
Now $n$ is the number of bond-operators $H_{D,ij}$ or $H_{O,ij}$  in the operator string
$S_K$ \cite{Sylj02}.  The MC procedure must
sample the space of all states $|\alpha\rangle$ and all operator
sequences $S_K$ with the relative weight
\begin{equation}
W(\alpha, S_K)=\frac{\beta^n(K-n)!}{K!}\langle\alpha | S_K |\alpha\rangle . \label{weight}
\end{equation}
Denoting a propagated state by
\begin{equation}
|\alpha(p)\rangle=\prod_{i=1}^pH_i|\alpha\rangle,
\end{equation}
the matrix element in Eq.(\ref{taylor}) can be written as a
product of elements with the form
$\langle\alpha(p)|H_{ij}|\alpha(p-1)\rangle$, which is equivalent to 
$$\langle S^z_i(p) L^z_i(p) S'^z_j(p)| H_{ij}| S^z_i(p-1) L^z_i(p\!-\!1) S'^z_j(p\!-\!1)\rangle. $$
We shall refer to these matrix elements as ``vertices.''   The matrix element in Eq.(\ref{taylor}) can be
viewed as a list of such vertices.

In the operator-loop algorithm, two basic updates ensure that the
complete SSE space is sampled.  The diagonal update attempts to
exchange diagonal operators $H_{D,ij}$ with unit operators $I$.  The
probability for inserting a diagonal operator (exchanging it for a
unit operator) at position $p$ in the operator sequence is
\begin{equation}
P_{\text{insert}}=\frac{N_p\beta \langle\alpha(p)|
H_{D,ij}|\alpha(p)\rangle}{K-n},
\label{eq:ins}
\end{equation}
while the probability for removing a diagonal operator is
\begin{equation}
P_{\text{remove}}=\frac{K-n+1}{N_p\beta \langle\alpha(p)|
H_{D,ij}|\alpha(p)\rangle}.
\label{eq:rem}
\end{equation}
The total number of nearest-neighbor pairs on the lattice is denoted $N_p$. 
In a diagonal update, one exchange attempt is made for each diagonal
and unit operator.

The second type of update is a global operator-loop update, which
leaves unit operators unaffected.  This update forms
and flips a closed loop of spins in the vertex list.  In this process,
both the affected vertices and states are changed.  For a detailed description of the operator-loop move,
we refer to Ref.\cite{Sylj02}.  In the absence of spin-orbit coupling, the operator-loop
update together with the above diagonal update ensure that the
complete SSE configuration space is sampled.  But when spin-orbit coupling is included,
the Ising-like orbital angular momentum $L_i^z$ must also be updated. 
Due to the classical nature of $L_i^z$, no terms in the Hamiltonian are able to flip $L_i^z$ and the 
orbital angular momenta must be updated in a separate move, which we describe next.

Additional flips of the spin variables $S_i^z$ and $S_j^{\prime z}$ are allowed in case 
no string operator acts on sites $i$ or $j$.  The weight of the configuration then
remains unchanged after a spin flip. The same is true for the orbital angular momenta $L_i^z$:
if no string operator acts on a given variable $L^z_i$, it can be flipped with no associated change in weight. 

This method must be modified at or below $\TC $, where 
many string operators act on a given $L^z_i$.  Flipping $L^z_i$ then causes the corresponding vertices to change
with the associated weight change given by Eq.(\ref{weight}), which must be must be taken into account 
when updating the orbital angular momenta.  Denoting the weight of a vertex by
\begin{equation} 
W_p= \langle\alpha(p)|H_p|\alpha(p-1)\rangle,
\end{equation} 
then the acceptance probability of flipping an orbital momentum is
\begin{equation}
P=\frac{\prod_i^{I}W'_{p(i)}}{\prod_i^{I}W_{p(i)}}, 
\end{equation}
where  $W'$ indicates the weight after the flip and $W$ the weight before the flip. 
The product runs over all vertices containing the orbital momentum $L_i^z$. 

The method described here works well over the range of parameters studied in this paper. 
It acts like a hybrid quantum-classical MC method where the Heisenberg spins are updated 
using a quantum MC loop algorithm but the Ising-like orbital angular momenta are updated 
separately using a classical single spin-flip MC method.  The effective field experienced by 
each orbital angular momentum depends on the quantum state $\{ |\alpha\rangle, S_K \}$ of the system. 
Since the orbital angular momenta saturate before the Heisenberg spins, it may be preferable 
to turn off the classical update at low temperatures in order to prevent freezing   
in a state different from the ground state.  It is straightforward to generalize the present 
method to include orbital angular momenta on both magnetic sublattices \cite{Reis08}, which
would allow the study of many other families of bimetallic oxalates.  Like the standard 
stochastic series method, this method can also easily be generalized to arbitrary spin size.

\section{Transition Temperature and Phase Diagram}

We begin our discussion by considering the lattice and boundary conditions. The honeycomb 
lattice consists of two sites per unit cell.  Using periodic boundary conditions on a two-dimensional  
lattice with $N\times N$ unit cells, the number of sites $N_s = 2N\times N$ varies between 32 ($N=4$) and 
32768 ($N=128$).  To study much larger system sizes would most likely require improvements in the algorithm.  

The thermodynamic expectation value of the sublattice magnetization 
on the Fe(II) sites is given by
\begin{equation}
\langle |S^z|\rangle= \frac{2}{N_s} \Bigl\langle \Bigl\vert \sum_i S_i^z \Bigr\vert \Bigr\rangle,
\end{equation}
where the summation runs over all $N_s/2$ sites.  In a similar manner, we define the sublattice magnetization on the 
Fe(III) sites, $\langle |S'^z|\rangle$, and the orbital angular momentum at the Fe(II) sites, $\langle |L^z|\rangle$. 
The absolute value is measured because the strict statistical average vanishes 
for finite systems due to time-reversal symmetry in zero magnetic field.  
The total magnetization is given by 
\begin{equation}
M=2\langle |S'^z|\rangle- 2\langle |S^z|\rangle - \langle |L^z|\rangle, \label{mag}
\end{equation} 
where we take $g=2$.  Recall that $\langle S^z \rangle $ and $\langle L^z\rangle $ have the same
sign because $\lo $ is negative.

\begin{figure}
\resizebox{\hsize}{!}{\includegraphics{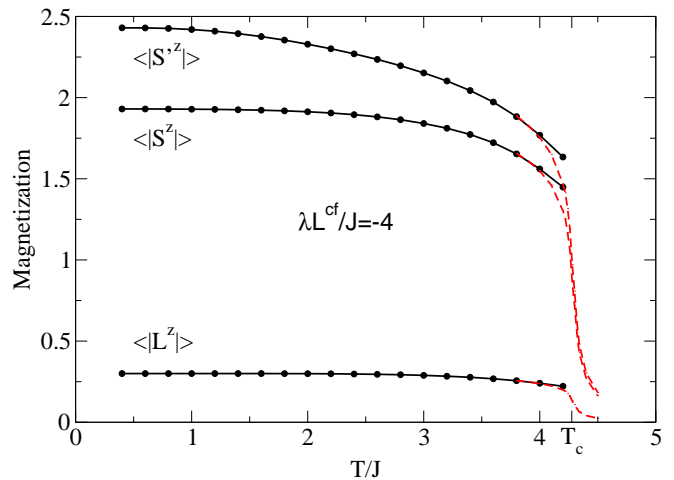}}
\caption{Sublattice magnetizations $\langle |S'|\rangle$,  $\langle |S|\rangle$ and $\langle |L^z|\rangle$ as a function of 
temperature for $\lo =-13.3 J$ and $\ld =0.3$.  The solid and dashed lines denote results for $4\times 4$ and $64\times 64$ 
unit cells, respectively. The critical temperature is $\TC =4.28 J$}
\label{magssl} 
\end{figure}

Sublattice magnetization curves for $\lambda=-13.3\, J$ and $\ld =0.3$ are plotted versus temperature 
in Fig.\ref{magssl}.  As shown later, these parameter values should be close to those expected for bimetallic 
oxalates that exhibit magnetic compensation.  Many features characteristic 
of the general magnetization curves appear in this figure.  At low temperatures, the results for different system 
sizes have converged to the thermodynamic limit.  But due to the diverging correlation length, 
increasing finite-size effects prevent an accurate determination of the critical temperature.  For the remainder
of this paper, we only present those portions of the magnetization curves that have converged in system size. 

At low temperatures, we can clearly see the effects of quantum fluctuations.  Since the spins are treated as
Heisenberg operators, the classical ground state is not an eigenstate of the Hamiltonian.  
The ground state magnetizations for the $S=2$ and $S'=5/2$ sublattices are 1.91 and 2.41, respectively, or about 
4$\%$ below the classical values.  These deviations are small because of the relatively high spin values. 
The figure also illustrates the physical mechanism responsible for magnetic compensation.  Due to the 
effective anisotropy induced by the spin-orbit coupling to the Ising-like orbital angular momentum, the 
$S=2$ sublattice magnetizes faster than the $S'=5/2$ sublattice. 
At $T=2$ K, the $S$ sublattice has almost reached its saturation magnetization but the $S'$ sublattice has not.  

\begin{figure}
\resizebox{\hsize}{!}{\includegraphics{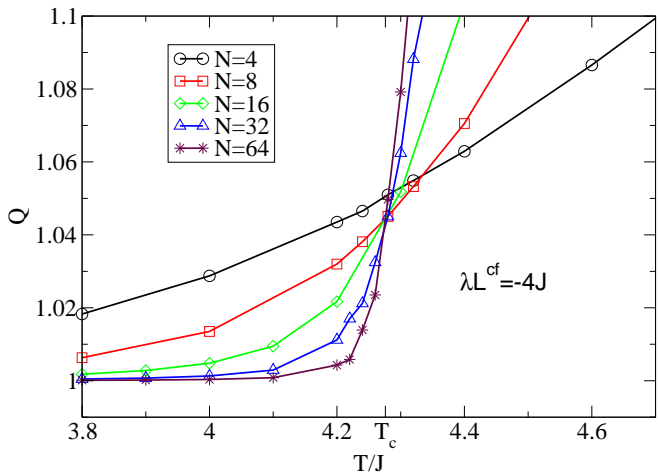}}
\caption{Binder ratio as a function of  temperature for $\lo \ld =-4 J$. Results for different system sizes intersect at the critical 
temperature $\TC =4.28 J$ The number of sites varies from 32 ($N=4$) to 8192 ($N=64$). }
\label{binder} 
\end{figure}

In order to accurately determine the critical temperature, we have measured the Binder ratio \cite{Bind81} of the magnetization,
\begin{equation} 
Q=\frac{\langle (S^z)^4\rangle^\frac{1}{4}}{\langle (S^z)^2\rangle^\frac{1}{2}},
\end{equation}
which is the ratio of two moments of the order parameter.  The Binder ratio should become size-independent at a second-order 
phase transition.  After plotting the ratio for various system sizes, the point of intersection gives the critical temperature. 
In Fig.\ref{binder}, we plot the Binder ration for $\lo \ld =-4J$, which is the value used in Fig.\ref{magssl}.
Finite size effects are clearly visible:  the ratios for system sizes $4\times 4$ and $8\times 8$ intersect at   
$T=4.34 J$ while the ratios for larger system sizes intersect at $T=4.28 J$. 
The convergence in system size improves with increasing anisotropy.

\begin{figure}
\resizebox{\hsize}{!}{\includegraphics{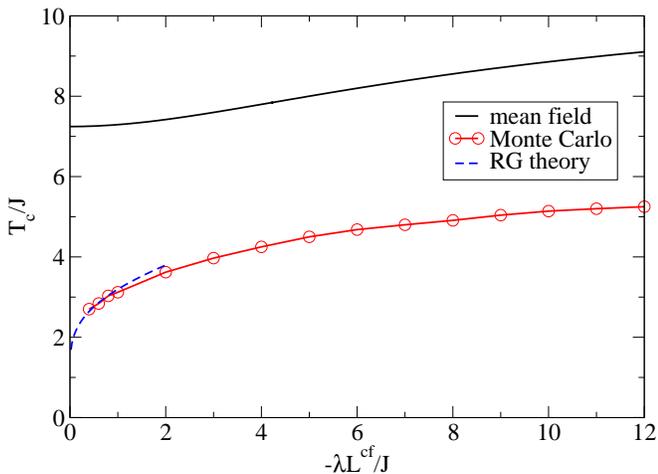}}
\caption{Critical temperature as a function of spin-orbit coupling $-\lo \ld $. Results from MF theory and quantum MC 
calculations along with the asymptotic behavior for small anisotropy from RG theory. }
\label{tcmfqmc} 
\end{figure}

The critical temperature depends only on the product $- \lo \ld $ and we compare
the MF and MC results for $\TC $ in Fig.\ref{tcmfqmc}.  According to the Mermin-Wagner 
theorem \cite{Merm66}, the critical temperature of a two-dimensional system must approach zero as 
the anisotropy vanishes.  One of the main weaknesses of the MF solution plotted in Fig.\ref{tcmfqmc}
is that it overestimates $\TC $ and predicts a finite critical temperature even for an isotropic system
with $-\lo \ld =0$.  By contrast, the MC results in Fig.\ref{tcmfqmc} indicate that 
$\TC $ rises very rapidly with small $-\lo \ld $.  However, even a small anisotropy induces  a 
critical temperature of order $J$.  In Fig.\ref{tcmfqmc}, we model the rapid initial rise in the 
critical temperature with the functional form 
\begin{equation}
\TC =\frac{a}{\ln (b /\vert \lo \vert )}.
\end{equation}
This dependency was first reported by a renormalization-group (RG) study \cite{Band88}, where the parameters $a$ and $b$ 
were related to the critical temperature of the three-dimensional isotropic model and the the anisotropy parameter, respectively. 
Here, we simply treat $a$ and $b$ as fitting parameters to demonstrate that our MC results are reasonable. 
Notice that MF theory overestimates $\TC $ by about 60\% in almost the whole parameter range of interest, 
an effect that must be taken into account when applying MF theory. 

\begin{figure}
\resizebox{\hsize}{!}{\includegraphics{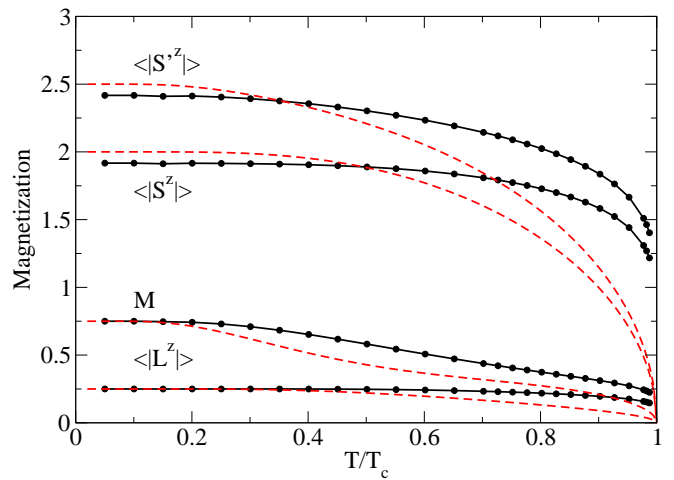}}
\caption{Total and sublattice magnetization calculated in MF theory (dashed line) and in the MC method (solid line) for 
$\lo =-12 J$ and $\ld =0.25$. We display only MC data that have converged in system size, therefore not quite reaching the 
critical temperature.}
\label{magmfqmc} 
\end{figure}

Scaling the temperature by $\TC$, we compare the MC and MF results for the total and sublattice
magnetizations in Fig.\ref{magmfqmc} for $\lambda=-12 J$ and $\ld =0.25$.  
Below 0.5$\TC $, the classical variable $L^z$ is essentially in its fully-polarized classical ground state 
within both the MF and MC solutions.  For the Heisenberg variables $\vS $ and $\vS' $, 
quantum fluctuations again cause deviations from the classical ground state. 
In the critical region, the sublattice magnetizations are described by the functional form 
$(T-\TC )^{\beta}$.  Whereas the MF exponent is $\beta=1/2$, the Ising-like exponent $\beta=1/8$ describes the MC data. 
Consequently, MF theory underestimates both sublattice magnetizations as $\TC $ is approached. 
The agreement between the MF and MC results
for the total magnetization $M= 2\langle |S'^z|\rangle- 2\langle |S^z|\rangle - \langle |L^z|\rangle$
is  generally better than for the individual sublattice magnetizations.   
Since the total spin $S_{\rm tot}^z = \sum_i S_i^z+\sum_j S_j^{\prime z} $ commutes with $H$, 
quantum fluctuations do not effect the total magnetization at $T=0$ \cite{Iva02} and $M(0)=2(S'-S)-\ld $.  
At higher temperatures, the MF solution underestimates $M(T)$, which has the same critical exponent $\beta =1/8$ as the sublattice magnetizations.

\begin{figure}
\resizebox{\hsize}{!}{\includegraphics{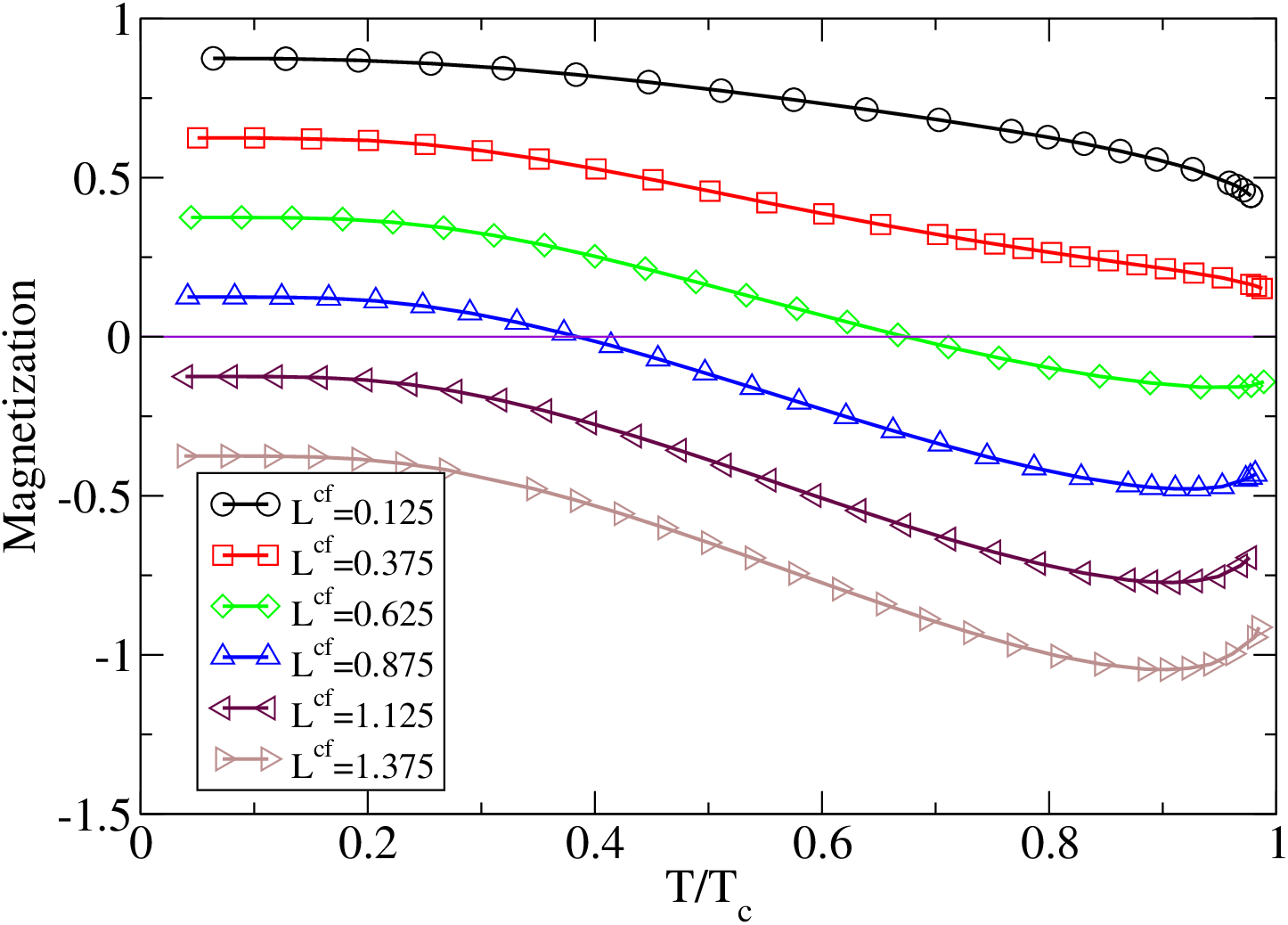}}
\caption{Total magnetization for $\lo =-8 J$ as a function of $\ld $ and temperature.}
\label{mag08} 
\end{figure}

Due to both the orbital contribution and the effects of anisotropy, the total magnetization
depends sensitively on $\ld $, as seen in Fig.\ref{mag08} for $\lo =-8 J$.  
When $\ld $ is small, $M$ is dominated by the $S'=2$ sublattice (defined to be positive) and it remains positive for all temperatures.  
When $\ld $ is large, $M$ is dominated by the $S$ sublattice and it remains negative for all temperatures. 
For intermediate values of $\ld $, there is competition between the two sublattices.  As shown, the  
magnetizations for $\ld =0.625$ and 0.875 exhibit a compensation point $\tcomp $ where $M(T)$ vanishes
due to a cancellation on the two sublattices. 

\begin{figure}
\resizebox{\hsize}{!}{\includegraphics{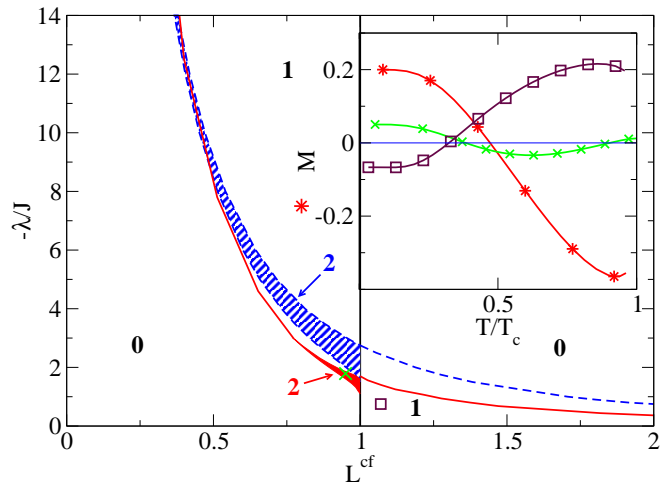}}
\caption{Phase diagram indicating the number of compensation points $\ncomp $.  The phase boundaries are 
given by the $\ld =1$ vertical line, as well as by the dashed (MF) and solid curves (MC).  
Two compensation points are found in the shaded regions. In the inset are shown the MC magnetization curves
for the three indicated points.}
\label{phase}
\end{figure}

By varying the temperature or some other control parameter such as strain \cite{Fish07}, the magnetization can be 
switched in the vicinity of a compensation point.  Due to the potential applications of this effect,
one main goal of this study was to determine the values of $\lo $ and $\ld $ where magnetic compensation may occur.
Our main result is the phase diagram of Fig.\ref{phase} denoting the number $\ncomp $ of compensation points
as a function of $\ld $ and $\lo $.  This phase diagram was previously 
determined using MF theory \cite{Fish07}.  In Fig.\ref{phase}, we display the MC results on top of the MF phase diagram. 
Since the magnetization of the classical ground state is given by $M=1-\ld $, $\ld =1$ is an important dividing line: 
for $\ld <1$,Êthe $S'=5/2$ sublattice dominates and the ground state magnetization is positive;   for $\ld >1$, the $S=2$ sublattice dominates
and the ground state magnetization is negative.   When $\ld < 1$ and $-\lo $ is large enough, the $S$ sublattice may magnetize faster 
than the $S'$ sublattice, producing a compensation point $\tcomp $ as the temperature is lowered. 
Similarly, when $\ld >1$ and $-\lo $ is sufficiently small, the $S'$ sublattice may magnetize faster than the
$S$ sublattice, again causing the two sublattice magnetizations to cancel at $\tcomp <  \TC $.  

The MF phase diagram contains an interesting region with not one but two compensation points.  
From Fig.\ref{phase}, it is evident that this area has shrunk and shifted, but remains finite despite the effect of
fluctuations.  It is impossible to say whether the MC region with $\ncomp =2$ survives in a narrow neck for all
$-\lo /J $ greater than about 1.1.  But for $-\lo /J  \ge 4.5$, there no discernible neck of $\ld $ where two 
compensation points can be found.  So it came as a surprise when a recently-studied Fe(II)Fe(III) bimetallic oxalate seemed, 
at first sight, to exhibit two compensation temperatures \cite{Tang07}.  However, those measurements 
are more naturally explained by an inverse Jahn-Teller transition, above which the $C_3$ symmetry 
of the lattice is violated \cite{Fish08}.  

For $\ld < 1$, the MF curve separating the $\ncomp =2$ and 1 region lies quite close to the MC curve 
separating the $\ncomp =0$ and 1 region when $-\lo /J$ is large.  These two curves appear to cross at about 
$-\lo /J \approx 12$.  For $\ld > 1$, the difference between the MF and MC regions
with $\ncomp = 1$ is much more pronounced, with the MC region about half the
size of the MF region.

In the inset to Fig.\ref{phase}, we show the magnetization curves for three selected points in the phase diagram.  These
points display either one ($\ld > 1$ or $\ld < 1$) or two compensation points.   Notice that the $S=2$ sublattice
dominates at low temperatures for the curve showing one compensation point with $\ld > 1$.

Using the experimental results, we may now estimate the parameter values that are relevant for the Fe(II)Fe(III) bimetallic oxalates.  
With $\lo =-147$ K \cite{Bleaney53} and $\TC = 45$ K, we find that $|\lo |/\TC \approx 3.3$.  The 
bimetallic oxalates should lie quite close to the phase boundary in Fig.\ref{phase} because some exhibit a compensation 
point but others do not \cite{comp}.  In Fig.\ref{magssl}, we used the parameter values $\lo =-13.3 J$ and $\ld =0.3$, 
corresponding to $\TC =4.24 J$ and $|\lo |/\TC \approx 3.13$, which is close to the expected ratio.  Since this point lies near
the phase boundary in Fig.\ref{phase}, it represents a reasonable estimate for the bimetallic oxalates.
These estimates yield an exchange constant of $J \approx 10.6$ K, about twice the MF estimate \cite{Fish07}.

\section{Conclusion}

We have studied a model for a molecule-based magnet using a hybrid quantum-classical 
MC method.  Our main result is the phase diagram in Fig.\ref{phase}, which displays the number of compensation 
points as a function of the spin-orbit coupling and orbital angular momentum.  
The region with two compensation points, previously found in MF studies \cite{Fish07}, has shrunk but remains finite.

With a simple scaling of the transition temperature, almost all MF results are qualitatively recovered in this MC study.
The surprising robustness of the MF results to the effects of fluctuations arises from two factors.  
First, the spin values $S=2$ and $S'=5/2$ in the Fe(II)Fe(III) bimetallic oxalates are high enough that 
quantum fluctuations are relatively unimportant.  Second, the important magnetic properties
depend on the difference between the magnetic moments on the two sublattices.  That difference is more immune
to the effect of fluctuations than the individual sublattice moments.  

Several of the discrepancies between the earlier MF results and experiments are resolved by our MC results.  
Within MF theory, the Curie-Weiss temperature $\Theta $ obtained from the high-temperature susceptibility 
$\chi \approx C/(T-\Theta )$ is always smaller in magnitude than $\TC $ \cite{Fish07} whereas experimentally 
it can be twice as large \cite{comp, Coronado00}.  While $\vert \Theta \vert $ should remain close to its MF 
value even in the presence of fluctuations, $\TC $ is reduced by about half from its MF value.  
Therefore, the ratio $\TC /\vert \Theta \vert $ is also suppressed by about 50\% due to fluctuations.  

The MC results for $\TC $ in Fig.\ref{tcmfqmc} also explain why bimetallic oxalates that exhibit magnetic
compensation tend to have transition temperatures roughly 10 K higher than those that do not \cite{comp}.  
If materials with magnetic compensation have $- \lo \ld \sim 4 J$, then Fig.\ref{tcmfqmc} 
suggests that materials without magnetic compensation will have $- \lo \ld \sim 2 J$ with $\TC $ reduced
by about 25\%.  By contrast, the MF result for $\TC $ in Fig.\ref{tcmfqmc} is relatively insensitive to $-\lo \ld $ and
cannot explain this sizeable suppression of the transition temperature.
  
Since our model Hamiltonian does not contain any interlayer couplings, $\TC $
vanishes if the singlet lies below the doublets in the $L=2$ Fe(II) multiplet.  Therefore, it seems
likely that one of the doublet always lies lowest in energy even in compounds that do not exhibit
magnetic compensation.  Of course, neglecting the interlayer coupling was just a convenient 
approximation within the current treatment.  While the relative insensitivity of the transition temperature 
to the interlayer separation and to the presence of radical spin-1/2 cations between the 
layers \cite{Coronado00, Clem97} suggests that the interlayer coupling is small, it must be 
present to support long-range magnetic order along the out-of-plane direction in zero magnetic field.

The present method can be easily generalized to consider spins of any size and spin-orbit coupling
on both magnetic sublattices.  Since quantum fluctuations grow with decreasing spin, the hybrid classical-quantum MC
technique may be particularly useful when considering low-spin bimetallic oxalates such as Ni(II)Mn(III) with $S=1$ and $S'=2$.
This would enable us to re-evaluate the magnetic phase diagrams obtained by Reis {\em et al.} \cite{Reis08}, where
compensation was found in ranges of $L$ and $L'$ but quantum and thermal fluctuations were not considered.
In order to improve the performance of the method, particularly close to $\TC $, one could use the 
directed-loop method \cite{Sylj02} to minimize the backtracking process in the loop construction.  Furthermore, the 
classical single-spin flip update could be augmented by a cluster update \cite{Sand03}.
We hope that the hybrid quantum-classical MC technique developed in this paper also proves of value in studies 
of other systems with both quantum and classical degrees of freedom.

\begin{acknowledgments}

We would like to acknowledge helpful conversations with Drs. Fernando Reboredo and Anders Sandvik.
P.H. acknowledges support by the Swedish Research Council. We are  grateful for the generous time allocation 
on the Ferlin cluster managed by  the Center for Parallel Computers at KTH.  R.F. acknowledges support by the Laboratory 
Directed Research and Development Program of Oak Ridge National Laboratory,
managed by UT-Battelle, LLC for the U. S. Department of Energy
under Contract No. DE-AC05-00OR22725 and by the Division of Materials Science
and Engineering of the U.S. DOE.

\end{acknowledgments}

\end{document}